\documentstyle[bibnorm,psfig]{lamuphys}

\begin{document}

\title{Force and Motion Generation of Molecular Motors:
A Generic Description}
\author{Frank J\"ulicher}
\institute{Institut Curie, Physicochimie Curie, 
UMR CNRS/IC 168, 26 rue d'Ulm,
75248 Paris Cedex 05, France}
\maketitle
\begin{abstract}
We review the properties of biological motor proteins which move along
linear filaments that are polar and periodic. The physics of the
operation of such motors can be described by simple stochastic models
which are coupled to a chemical reaction. We analyze the essential
features of force and motion generation and discuss the general
properties of single motors in the framework of two-state models.
Systems which contain large numbers of motors such as muscles and
flagella motivate the study of many interacting motors within the
framework of simple models. In this case, collective effects can lead
to new types of behaviors such as dynamic instabilities of the steady
states and oscillatory motion.
\end{abstract}
\section{Introduction}

An essential and striking features of living cells is their ability to
generate motion and forces. Important examples are cell
motility, the contraction of muscles but also active phenomena within
cells that allow for material transport and the motion of organelles,
for example during cell division and mitose. These movements
and forces are generated on the molecular level by protein molecules
that are driven by chemical reactions in a far from equilibrium
situation.  Important examples are motor proteins, enzymes which are
specialized to work as motors. In eucariotic cells, several families
of motor proteins exist which interact with the cytoskeleton, a
complex three-dimensional elastic network of long rod-like filaments
inside the cell which is essential for the mechanical
stability and integrity of the cell \cite{albe94}. 

\begin{figure}
\centerline{\psfig{figure=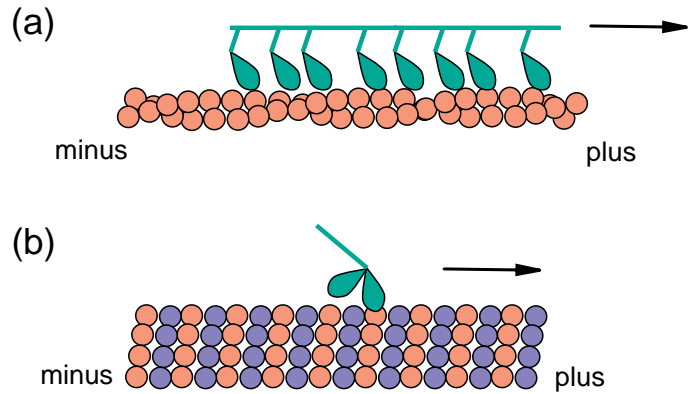,width=9cm}}
\caption{Schematic representation of molecular motors and track
filaments. (a) myosin interacting with actin
filaments. (b) kinesin moving along a microtubule. Both types of
filaments are polar and periodic, their two different ends are denoted
``plus'' and ``minus''.}
\label{f:motfil}
\end{figure}

A motor protein of the cytoskeleton interacts specifically with a
certain type of filament along which it is able to move in presence of
Adenosine\-tri\-phos\-phate (ATP) which is a chemical fuel. The
filaments serve as guides or tracks for the motion. Two types of
filaments play this role: microtubules and actin filaments. Both are
formed by a polymerization process from identical monomers (actin and
tubulin monomers, respectively), leading to a regular and periodic
structure. An important feature is their polarity: The filaments are
asymmetric with respect to their two ends. This symmetry has its
origin in the asymmetry of the monomers which form a polar filament
structure with two different ends which are denoted ``plus end'' and
``minus end''.  This polar symmetry is essential for motor operation
as it defines the direction of motion.  Motor proteins are classified
into several families: myosins, kinesins and dyneins. Myosins move
always along actin filaments and towards the plus end. Kinesins and
dyneins move along microtubules, kinesins move towards the plus end
and dyneins towards the minus end, see Fig. \ref{f:motfil}.

Myosins are prominent for their role in the contraction of muscles
\cite{huxl57,huxl69}.
In this case, many myosin molecules form a linear structure, a myosin
filament, which interacts with actin filaments arranged in
parallel. The action of myosin motors then induces the relative
sliding of the two types of filaments.  In muscle cells, a very large
number of filaments is organized in a regular way which on a
macroscopic scale leads to the muscular contraction.  Myosins also
occur within normal cells where they play an important role for cell
motility and the organization of actin.  Kinesins occur in large
numbers in neurons, where they play a key role in transport of
vesicles along the axon towards the synapse. Both types of motors have
two identical heads of a size of about 10-20nm which are the
elementary force-generating elements as well as a tail which is used
to attach the motor to another structure \cite{krei93}.
\begin{figure}
\centerline{\psfig{figure=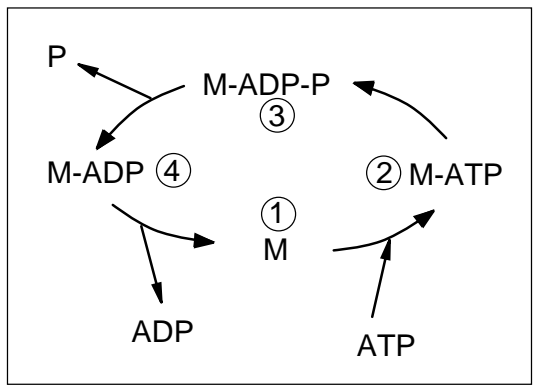,width=7cm}}
\caption{Chemical 
cycle of a motor molecule $M$. After completion of the cycle
on molecule ATP is hydrolyzed to
ADP and phosphate (P). }
\label{f:cycle}
\end{figure}

The energy source of this process is the hydrolysis reaction
$ATP\rightarrow ADP +P$ of ATP to
ADP and Phosphate (P). The motor protein $M$
(or more precisely, the head domain containing the ATP binding site)
undergoes a chemical cycle: it binds ATP and hydrolizes the bound
ATP. Subsequently it releases the products ADP and P. We denote the
different chemical states by M, M-ATP, M-ADP-P, and M-ADP,
respectively.  After completion of the cycle (M+ATP $\rightarrow$
M-ATP $\rightarrow$ M-ADP-P$ \rightarrow$ M-ADP+P $\rightarrow$
M+ADP+P) the motor is unchanged. However, during this process it has
undergone conformational changes and it has hydrolyzed one ATP
molecule, see Fig. \ref{f:cycle}.  The different conformations which
occur during the chemical cycle in general have different geometries
and properties and can in particular have different interaction
characteristics with respect to the filament \cite{huxl71}. As a
result, the motor protein undergoes chemistry-driven changes between
strongly and more weakly bound states (``attachments'' and
``detachments''). This coupling between chemistry and binding permits
the creation of motion along a polar filament
\cite{hill74,spud90,juli97b}.

The force and motion generation of individual or groups of motor
proteins can be studied experimentally by a variety of techniques. In
so-called motility assays, motors are attached to a substrate
\cite{kron86,hara87,ishi91,wink95}. Filaments in solution bind to the
motors and in presence of ATP start moving along the surface. In these
experiments typically several or many motors interact with a single
filament.

\begin{figure}
\centerline{\psfig{figure=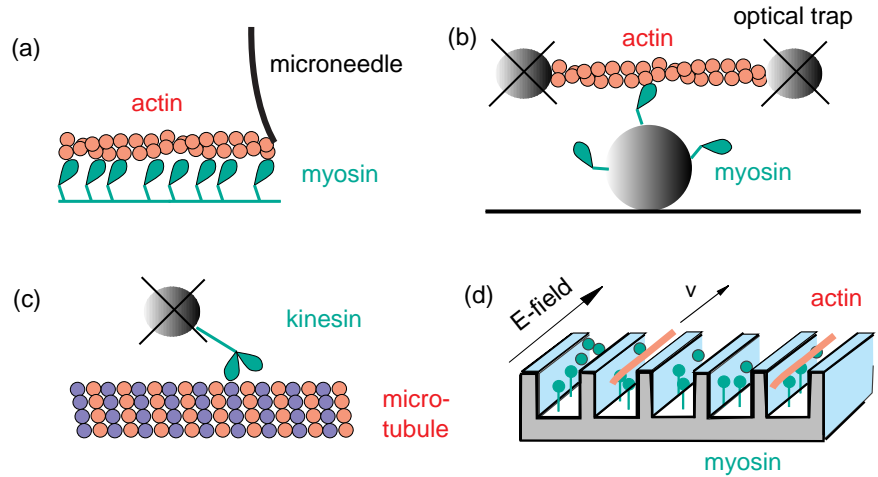,width=11.5cm}}
\caption{Examples for micro-manipulation experiments. (a) Forces induced
my myosin on actin can be measured by the deflection of a micro-needle
\protect\cite{ishi91}.
(b) Force measurement using optical traps \protect\cite{fine94}. (c)
Forces generated by a single kinesin molecule observed by
displacements of a bead in an optical trap \protect\cite{svob93}. (d)
Forces induced by an electric field $E$ in a motility assay using
linear grooves to orient the filaments \protect\cite{rive98}.}
\label{f:motexp}
\end{figure}

In order to observe the forces generated by an individual motor, the
processivity of the motor becomes important. Myosin is not processive.
During the chemical cycle it detaches from the filament during a
significant period of time. During this time it can easily diffuse
away from the filament if it is not held in place. Forces generated by
single myosin molecules have been observed by different techniques,
see Fig. \ref{f:motexp} (a) and (b).  Micron sized bead have been
coated with low density of motors and optical traps have been used to
bring a filament in contact with the bead and possibly only a single
motor \cite{fine94}.  Forces on the filament are then measured by observing
the displacement in the trap. Another possibility is to fix the actin
and to manipulate the bead with an optical trap. Such experiments
reveal that the motor induces stochastic displacements of the order of
$5-10$nm which last for several milliseconds and peak forces of the order
of $1$pN \cite{ishi91,fine94}.

Kinesin, is a processive motor \cite{bloc98}. A single kinesin moves
along microtubules for a distance of the order of micrometers before
losing the filament.  Motion of single kinesin molecules can therefore
be observed directly, either by attaching a small bead to the motor
and observing the displacement of the bead or by directly marking the
molecule with a fluorescent dye. Optical traps have been used to study
the velocity of motion as a function of an applied load
\cite{svob93}. It has been shown, that kinesin moves in a step-wise
fashion with characteristic steps of 8nm size \cite{schn97}.  This
step size coincides with the period of kinesin binding sites along
microtubules which demonstrates that the ATP-driven reaction cycle
induces steps between periodically spaced binding sites.  The
characteristic time-scales for kinesin motion are milliseconds and
forces up to 5pN are generated
\cite{svob93,hunt94}.

The standard explanation for the processivity of Kinesin is based on
the fact that each molecule consists of {\em two} head domains
which both hydrolize ATP and undergo the chemical cycle
in a coordinated way
\cite{bloc98,mand99}. In this situation, processivity is possible if both
heads very rarely detach from the filament at the same time thus
allowing the motor to keep attached while displacing. Single heads do
have the capability to generate forces and motion but in most cases
are not processive. Recent experiments however show evidence that
single kinesin heads can move processively in certain cases
\cite{okad99}.  Myosin molecules also often have two heads. However,
even with two heads they do not become processive since every myosin
head has a tendency under normal operating conditions to be unbound
during about 90\% of its chemical cycle.

From the point of view of a general classification of the physical
mechanisms which can lead to motion generation, motor proteins fall in
a class of systems characterized by the fact that they operate on {\em
molecular scales} and generate motion along a one-dimensional polar
and periodic structure (sometimes called ratchet \cite{feyn66}) by a
non-equilibrium {\em rectification process}
\cite{buet87,land88,ajda92,magn93,pros94,pesk94,astu94,doer95,astu97}. 
For a system on a molecular scale, fluctuations play an important role
for its function and properties.  These fluctuations can be both of
thermal origin or they can arise due to the stochasticity of
individual molecule chemical reactions. Therefore, a theoretical
description requires the use of concepts of non-equilibrium
statistical physics. Biological motor proteins are the most prominent
examples for these systems. However, there is a growing number of
artificially designed ``molecular motors''
\cite{rous94,fauc95a,gorr96,gorr98}, suggesting that the physics of these
systems is relevant for micro-and nano-technological devices.

In the subsequent sections, we describe a simple modelization of the
generic aspects of these systems, using biological motor proteins as a
guide. In section 2 we introduce the basic concepts of such a
modelization. Properties of single motor motion which follow from this
description are described in Sect. 3. In section 4, we discuss the
consequences of many motor systems and demonstrate that new phenomena
such as dynamical instabilities follow naturally from these
models. Finally, we present in section 5 a discussion and an outlook.

\section{Simplified models}

In order to keep the description simple and to focus on generic
properties, we do not aim to capture microscopic structural details of
biological motor proteins. We use a simplified picture taking into
account the periodicity and polarity of the linear track and the fact
that during the chemical cycle conformational changes occur
\cite{juli97b}.

The main assumption required for this simplification is a separation
of time-scales \cite{hill74,juli97b}. Even though a macromolecule like
a protein is characterized by a large number of microscopic degrees of
freedom, most of them relax on time scales shorter than the typical
relevant time scales for the chemical cycle. These degrees of freedom
are therefore to a good approximation thermally equilibrated. Only a
few slow degrees of freedom have to be described by dynamical
equations.  These relevant degrees of freedom are collective modes of
the system.  Examples are chemical reaction coordinates but also the
overall position variable of the motor with respect to the
filament. The latter is important in order to describe the coupling
between chemistry and motion.

\subsection{Energy landscapes and chemical transition rates}

Following Ref. \cite{hill74}, we assume for simplicity that the
chemistry of a single head can be described by a number $m$ of
discrete states or conformations $i=1..m$. Many biochemical models
focus on the four states M, M-ATP, M-ADP-P, M-ADP but other reaction
intermediates could also be included \cite{hill74,lymn71}.
Transitions between these states are fast compared to the typical time
between transitions and the mechanical action or motion of the motor.
Therefore we use a chemical kinetics description for changes between
states.

If we consider a motor in conformation $i$, we can define a potential
or interaction energy profile along the filament. Suppose that one
small region of the motor, e.g. in the tail, is used to transmit
forces or to attach a cargo.  We imagine this point to be held at a
position $x$ along the filament. We can now define $W_i(x)$ to be the
energy of the motor, including possibly bound ATP, ADP or P, and
including the energy of the filament as the motor is held at position
$x$. This total energy is in fact an effective free energy defined
formally by integrating over all rapidly relaxing microscopic degrees
of freedom but keeping the enzyme in its chemical state. The
conformation of the system motor-filament is then fully characterized
by the pair $\{i,x\}$ of internal state and position with respect to
the filament.  Note, that the potentials reflect the symmetry
properties of the filament. If the filament is polar and a periodic
array of identical monomers, the potentials are periodic with period
$l$, $W_i(x)=W_i(x+l)$ and asymmetric, $W_i(x)\neq W_i(-x)$.

In order to describe the dynamics of the system $\{i,x\}(t)$, we use a
stochastic overdamped dynamics at constant temperature $T$ 
within a given state $i$
\begin{equation}
\eta_i \frac{d}{dt} x = - \partial_x W_i(x) + \zeta_i(t) \quad .
\label{eq:langevin}
\end{equation}
Here $\eta_i$ is a protein friction and $\zeta(t)$ is a Gaussian white
noise in state $i$ with $<\zeta_i(t)\zeta_j(0)>=2\eta_i\delta(t)$.  The
chemical reactions between states $\{i,x\}$ and $\{j,x\}$
\begin{eqnarray}
& \omega_{ij}(x) &\nonumber \\
\{i,x\} & \rightleftharpoons & \{j,x\}\label{eq:trans} \\ 
& \omega_{ji}(x) & \nonumber
\end{eqnarray}
occur with Poisson statistics with reaction rates $\omega_{ij}(x)$.
Since the position variable $x$ is also a conformational degree of
freedom (the motor in general changes its shape while displacing),
transition rates are in general $x$-dependent.  Note, that for
simplicity in Eq. (\ref{eq:trans}) we have assumed that transitions
between states happen instantaneously and without displacement.
Furthermore, we have used the fact that thermal relaxation is very
fast compared to the chemical cycle and all rapid degrees of freedom
are equilibrated at constant temperature $T$. In fact, the typical
relaxation time of temperature gradients which have developed on a
length scale $l$ can be estimated as $\tau=Cl^2/\kappa$, where $C$ is
the specific heat of the material per volume and $\kappa$ the thermal
conductivity. Using typical values for water and length scales of the
order of $10$nm we find $\tau \simeq 10^{-6}-10^{-8}$s, which is fast
compared to typical cycle times of several ms. This argument shows
that the motor operates isothermally, i.e. temperature gradients are
not created and cannot be used to generate motion as e.g. in the case
of Feynman's ratchet \cite{feyn66}.

It is now convenient to use a Fokker-Planck description \cite{risk84}
and to
introduce distribution functions $P_i(x,t)$ for the probability to
find within an ensemble of identical systems the motor at time $t$ at
position $x$ in state $i$. These distributions then obey the equations
\begin{eqnarray}
\partial_t P_i + \partial_x J_i & = & \sum_{j\neq i} 
(\omega_{ji}(x) P_i(x)-\omega_{ij}(x) P_j(x) ) \\
J_i & = & \eta_i^{-1}(-k_B T \partial_x P_i-P_i 
\partial_x W_i + P_i f_{\rm ext})
\quad .\label{eq:FP}
\end{eqnarray}
The total density and total current
\begin{eqnarray}
P(x,t)&=&\sum_{i=1}^m P_i \\
J(x,t)&=&\sum_{i=1}^m J_i 
\end{eqnarray}
obey the conservation law $\partial_t P+\partial_x J=0$. The average
velocity in the steady state with stationary and periodic distribution
function $P_i(x)=P_i(x+l)$, $\partial_t P_i=0$ is given by
\begin{equation}
v= \int_0^l J dx/\int_0^l P dx
\end{equation}

In order to characterize the chemical rates, we first introduce the
chemical potentials of the fuel and hydrolysis products in bulk
solution. We denote $\mu_{ATP}$, $\mu_{ADP}$ and $\mu_P$ the free
energy per ATP, ADP or P molecule, respectively. As an illustrative
example, we first consider the four chemical states M ($i=1$), M-ATP
($i=2$), M-ADP-P ($i=3$) and M-ADP ($i=4$), often encountered for
biological motor proteins. A general reaction kinetics for all eight
reaction rates which is consistent with the ATP hydrolysis reaction
can be written by using four different kinetic coefficients:
\begin{equation}
\begin{array}{l l}
\omega_{12}=\alpha_{1}\exp[(W_1+\mu_{ATP})/k_B T] & 
\omega_{21}=\alpha_{1}\exp[W_2/k_B T] \\
\omega_{23}=\alpha_{2}\exp[W_2/k_B T] & 
\omega_{32}=\alpha_{2}\exp[W_3/k_B T] \\
\omega_{34}=\alpha_{3}\exp[W_3/k_B T] & 
\omega_{43}=\alpha_{3}\exp[(W_4+\mu_P)/k_B T] \\
\omega_{41}=\alpha_{4}\exp[(W_4+\mu_{P})/k_B T] & 
\omega_{14}=\alpha_{4}\exp[(W_1+\mu_{ADP}+\mu_P)/k_B T] 
\end{array} \quad , \label{eq:cycle}
\end{equation}
Here, we have used the condition of detailed balance of the rates.
The functions $\alpha_i(x)$ characterize the possible reaction
scenarios.  Note, that since transitions are fast and therefore occur
for fixed $x$, the chemical rates do not depend on the external force
$f_{\rm ext}$ or local stresses. Force-dependent are only
displacements described by equation (\ref{eq:langevin}). The present
modelization differs in this respect from models which use discrete
transitions also to describe displacements. In this case 
chemical rates are strain dependent \cite{hill74,leib93,duke96}.
From a physical point of view, all models are of course equivalent.

\subsection{Two-state models}

It is useful to further simplify the generic description introduced
above. The m-state model allows in principle to describe many details
of the chemical cycle and the various conformations of the
motor. However, it contains a large number of free parameters which
are unknown. It is therefore useful to further simplify the model.  In
fact, in order to describe physical aspects of motion generation and
force generation, it is sufficient to keep only two different
states \cite{ajda92,pros94,pesk94}. 

\begin{figure}
\centerline{\psfig{figure=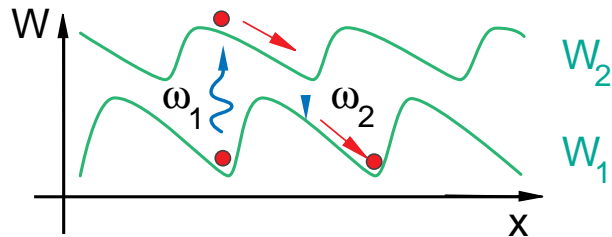,width=8cm}}
\caption{Two state model defined by two polar and periodic potentials 
$W_1$ and $W_2$ as well as periodic transition rates $\omega_1$ and
$\omega_2$.  Pumping between the two states induces average motion.}
\label{f:twostatemod}
\end{figure}

We rewrite the Fokker-Planck equations (\ref{eq:FP})
for two states $i=1,2$:
\begin{eqnarray}
\partial_t P_1 + \partial_x J_1 &=& -\omega_1(x) P_1 + \omega_2(x) P_2 \nonumber \\
\partial_t P_2 + \partial_x J_2 &=&  \omega_1(x) P_1 - \omega_2(x) P_2 
\quad ,\label{eq:2s}
\end{eqnarray}
where we have introduced $\omega_1=\omega_{12}$ and
$\omega_2=\omega_{21}$ and the currents are the same as introduced in
Eq. (\ref{eq:FP}). This system is sketched in Fig.
\ref{f:twostatemod} for an example of shifted periodic
and asymmetric potentials.
\begin{figure}
\centerline{\psfig{figure=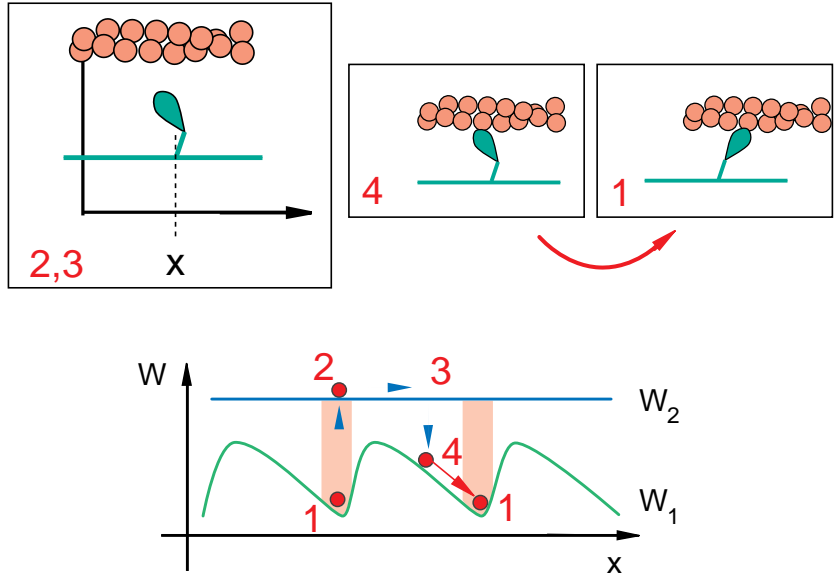,width=11cm}}
\caption{Two state model representing a situation motivated by the
functioning of myosin. (1) Binding an ATP molecule, myosin detaches
from actin (2).  After hydrolysis (3), it rebinds and generates a
force (4) and a displacement.  In a two state model, two potentials
$W_1$ and $W_2$ characterize attached and detached states with the
tail at position $x$. The shaded areas are ``active sites'' where
ATP-driven transitions occur.}
\label{f:myosin}
\end{figure}

This two-state model is still very flexible and allows to describe
situations which capture many of the physical aspects of biological
protein motors.  Fig. \ref{f:myosin} shows choice of potentials $W_1$
and $W_2$ adapted to the commonly accepted picture of myosin function
\cite{spud90}: a myosin head detaches from the actin filament after
binding ATP. In the unbound state ATP is hydrolyzed
(M-ATP$\rightarrow$ M-ADP-P).  The head (M-ADP-P) is now again able to
bind actin. As it encounters a binding site along the filament, it
re-attaches under phosphate release. After reattachment, a
force-generating step occurs and ADP is released, which completes the
chemical cycle. As illustrated in Fig. \ref{f:myosin} this process can
be captured by two different potentials, $W_1$ and $W_2$ representing
the unbound state (a flat potential) and the bound state (a potential
with periodic structure) respectively. After the force-generating step
(power-stroke), the $x$ variable has reached a potential minimum.
Here, the system can be actively excited to the unbound state under
ATP binding.  As it reattaches to the filament, the slope of the
potential reflects the mechanical force generated at this point. A
displacement is generated as the system slides downhill along the
energy profile to reach the potential minimum. Note, that
microscopically this displacement could correspond either to a tilt of
the head domain as sketched in the figure or to other more complex
processes.  The microscopic structure associated with this
displacement is not characterized by this description.

In a two-state picture, the chemical reaction cycle as described by
the kinetic equations (\ref{eq:cycle}) has to be divided into two
substeps.  One possibility is to introduce the forward and backward
rates $\alpha_1$ and $\alpha_2$ for the combined process of ATP-binding
and hydrolysis
\begin{equation}
\begin{array}{l l l}
 & \alpha_1 & \\
\hbox{\rm M + ATP} & \rightleftharpoons & \hbox{\rm M-ADP-P}  \\
 & \alpha_2 & 
\end{array}
\quad ,\label{eq:alpha}
\end{equation}
and the rates $\beta_1$ and $\beta_2$ which describe
the process of product release and binding:
\begin{equation}
\begin{array}{l l l}
 & \beta_1 & \\
\hbox{\rm M + ADP + P} & \rightleftharpoons &  \hbox{\rm M-ADP-P} \\
 & \beta_2 & 
\end{array}
\quad .\label{eq:beta}
\end{equation}
The complete chemical cycle is now the subsequent transitions
$\alpha_1$ and $\beta_2$. As long as $\alpha_2$ and $\beta_1$ are
nonzero, there is a nonvanishing probability for an inversion of the
cycle (i.e. ATP generation) by following the steps $\alpha_2$ and
$\beta_1$.  We define $W_1$ to be the energy of a free motor together
with the product molecules (M + ADP + P) and $W_2$ to be the energy of
the complex M-ADP-P. Detailed balance of the chemical reactions then
implies
\begin{eqnarray}
\frac{\alpha_1}{\alpha_2}& = &e^{(W_1-W_2+\Delta\mu)/k_BT}\\
\frac{\beta_1}{\beta_2} &=& e^{(W_1-W_2)/k_B T} \quad ,
\end{eqnarray}
where we have introduced the chemical driving force
\begin{equation}
\Delta\mu\equiv\mu_{ATP}-\mu_{ADP}-\mu_P \quad .\label{eq:dmu}
\end{equation}
The transition rates of the two-state model are the superpositions
$\omega_i=\alpha_i+\beta_i$. Introducing two unknown functions
$\alpha(x)$ and $\beta(x)$ which describe conformation dependent
energy barriers, we can therefore write
\begin{eqnarray}
\omega_1(x) & =& \alpha(x) e^{(W_1+\Delta\mu)/k_B T} 
+\beta(x) e^{W_1/k_BT} \nonumber \\
\omega_2(x) & =& [\alpha(x)+\beta(x)] e^{W_2/k_B T} \label{eq:omega_myosin}\quad .
\end{eqnarray}
Note, that other choices to divide the reaction cycle in two relevant
substeps leads to the same result, but redefines the arbitrary
functions $\alpha$ and $\beta$ and shifts the potential $W_2$ by a
constant value.

The functions $\alpha(x)$ and $\beta(x)$ define the coupling of the
chemical reaction to conformation. Very important is the concept of
localized or conformation-dependent transitions where the functions
are peaked within a narrow $x$-interval but negligible outside this
interval. An example is the ATP-binding step which in
Fig. \ref{f:myosin} is restricted to occur within an ``active region''
of conformation space corresponding to the potential minimum while the
conformations at the beginning of a force-generating power stroke are
not supposed to bind ATP. As we will describe in the subsequent
sections, the localization of transitions via the functions $\alpha$
and $\beta$ plays an important role for many interesting cases.

\begin{figure}
\centerline{\psfig{figure=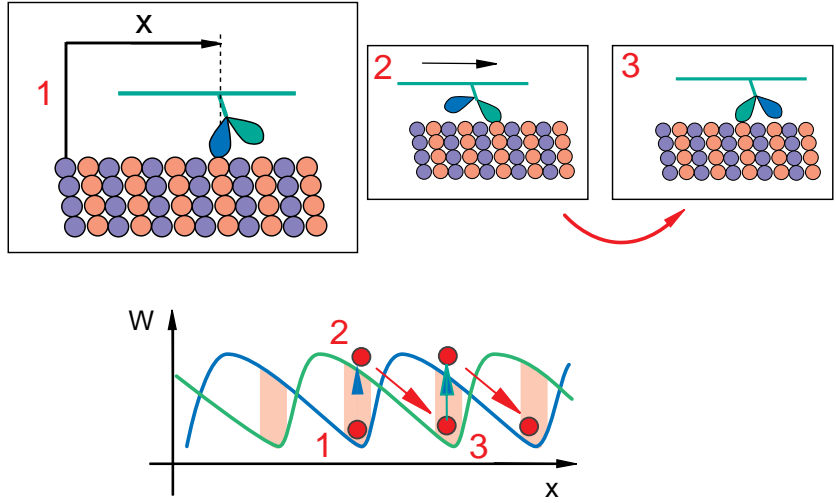,width=11cm}}
\caption{Hand over hand motion suggested for kinesin. At any time, one 
head is bound and the second head moves towards the next binding
site. This situation can be represented by describing both heads by
identical but shifted energy landscapes $W_1(x)$ and $W_2(x)$. }
\label{f:kinesin}
\end{figure}

Similar to the case of myosin, the two state model can also be adapted
to other situations such as the motion of kinesin molecules
containing two heads both of which contain an ATP binding site. In
principle, a general description would require eight internal states
and a complex reaction scenario.  A possible simplification arises
from the idea of a coordinated binding
and unbinding of the two heads in a hand-over-hand fashion as shown
schematically in Fig. \ref{f:kinesin}
\cite{bloc98}. Such a coupling would reduce the number of relevant
degrees of freedom \cite{pesk95}. In a two-state model,
this feature can be captured in a simplified way by associating each
state with one of the heads being bound to the filament. Denoting the
motor with head $1$ or head $2$ bound to the filament by $M_1$ and
$M_2$, respectively, we define the energies of these situations $\bar
W_1$ and $\bar W_2$.  Because the two heads forming a kinesin motor
are identical, the energy landscapes of both states are two identical
potential profiles which are shifted with respect to each other by one
monomer period $l/2$ on the filament: $\bar W_1(x)=\bar W_1(x+l)$;
$\bar W_2(x)=\bar W_1(x+l/2)$, see Fig. \ref{f:kinesin}. However, the
characteristic step-size of an individual head $l$ in this picture
corresponds to two monomer sizes. Therefore, the potentials $\bar W_i$
are $l$-periodic, while the total system is invariant under
$x\rightarrow x+l/2$ if at the same time the two states are
exchanged. Therefore the transition rates also obey
$\omega_1(x)=\omega_1(x+l)$ and $\omega_2(x)=\omega_1(x+l/2)$. Again,
the idea of active regions and localized transitions is
important. Assuming that all transitions occur at conformations which
correspond to the potential minimum, we obtain a system which operates
in an almost deterministic way where the chemical cycle is closely
correlated to a particular displacement. In the case of non-localized
transitions, the chemical cycle is related to motion in a more
irregular way.

A useful representation of the transition rates in the hand-over-hand
picture is to assume that a full ATP hydrolysis cycle changes $M_1$ to
$M_2$:
\begin{equation}
M_1 + ATP \rightleftharpoons M_2 + ADP +P \quad .
\end{equation}
Because of the symmetry between the two heads, the  reaction
$M_2 + ATP \rightleftharpoons M_1 +ADP +P$ occurs with the same
rates. This leads to the total transition rates
\begin{equation}
\begin{array}{l l l l l}
\omega_1(x) & = & e^{\bar W_1(x)/k_BT} [\bar\alpha(x) e^{\bar\Delta\mu/k_B T} 
&+& \bar\alpha(x+l/2)] \\
\omega_2(x) & = & e^{\bar W_2(x)/k_BT} [\bar\alpha(x)  
& +& \bar\alpha(x+l/2) e^{\bar\Delta\mu/k_BT\}}]
\end{array}
\quad . \label{eq:omega_kinesin}
\end{equation}
The unknown function $\bar\alpha(x)=\bar\alpha(x+l/2)$ is
$l$-periodic. Note, that the choice given in
Eq. (\ref{eq:omega_kinesin}) is a special case of
Eq. (\ref{eq:omega_myosin}) if we identity $\Delta\mu=2\bar\Delta\mu$,
$\alpha(x)=\bar\alpha(x)e^{-\bar\Delta\mu/ k_B T}$,
$\beta(x)=\bar\alpha(x+l)$, $W_2=\bar W_2+\bar\Delta\mu$ and
$W_1=\bar W_1$. This example demonstrates that
Eq. (\ref{eq:omega_myosin}) is a general choice which can describe
very different types of couplings of an ATP hydrolysis cycle to a two
state model.

We have now defined a two-state model which can describe a system
undergoing an ATP hydrolysis cycle and moving along a periodic
structure.  As discussed above, this model is sufficiently flexible to
be adapted to situations which resemble the widely accepted pictures
of myosin and kinesin functioning.

\section{Single motors}

We will now discuss general properties of the two-state model for a
single motor introduced above \cite{pros94,chau94,chau95}.  Two
generalized forces act on the system leading to an out-of equilibrium
situation. These are the chemical ``force'' $\Delta\mu$ introduced in
Eq. (\ref{eq:dmu}) and the mechanical force $f_{\rm ext}$. If both
generalized forces are kept constant, the system eventually attains a
steady state with $\partial_t P_i=0$. The steady state distribution
functions satisfy two coupled differential equations of second order
\begin{eqnarray}
k_B T \partial_x^2 P_1 + (\partial_x P_1)( \partial_x W_1-f_{\rm ext})
-P_1\partial^2_x W_1
&=& \eta(\omega_1 P_1 - \omega_2 P_2)\nonumber \\
k_B T \partial_x^2 P_2 + (\partial_x P_2)( \partial_x W_2-f_{\rm ext})
-P_2\partial^2_x W_2 
&=& -\eta(\omega_1 P_1 - \omega_2 P_2) \quad , \label{eq:steadystate}
\end{eqnarray}
where we have for simplicity assumes that the friction $\eta$ is the
same for both states. This set of equations together with periodic
boundary conditions $P_i(x)=P_i(x+l)$ defines the steady state
distributions.  They can be calculated in special cases analytically,
but in general numerical integration techniques are used. For each
pair $(\Delta\mu,f_{\rm ext})$, there is a uniquely defined average
velocity
\begin{equation}
v = \frac{1}{\eta}\int_0^l dx [P_1 (f_{\rm ext} - \partial_x W_1) 
  + P_2 (f_{\rm ext} - \partial_x W_2)] \quad , \label{eq:vss}
\end{equation}
where the $P_i$ satisfy the normalization condition
\begin{equation}
\int_0^l dx [P_1+P_2]=1 \quad .\label{eq:norm2s}
\end{equation}
Similarly, we can introduce the ATP
hydrolysis rate $r$ which denotes the number of chemical cycles
performed per unit time \cite{juli97b}. Using the rates introduced in
Eqns. (\ref{eq:alpha}) and (\ref{eq:beta}),
\begin{equation}
r = \int_0^l dx[\alpha_1 P_1-\alpha_2 P_2] = \int_0^l dx [
\beta_2 P_2 -\beta_1 P_1] \quad .\label{eq:rss}
\end{equation}
If $\Delta\mu=0$,
the transition rates defined by Eq. (\ref{eq:omega_myosin})
satisfy
\begin{equation}
\omega_1/\omega_2=e^{-\Delta W/k_BT} \quad , \label{eq:det_bal}
\end{equation} 
where
\begin{equation}
\Delta W(x) = W_2(x)-W_1(x) \quad .
\end{equation}
This condition of detailed balance for the total transition rates
indicates that transitions are just thermal fluctuations and that the
system is not driven chemically.  If the external force also vanishes,
the steady state is a thermal equilibrium with $P_i=N
e^{-W_i(x)/k_BT}$ for which $v=0$ and $r=0$.  For $\Delta\mu>0$, the
system is chemically driven. If no external force is applied
spontaneous motion with $v\neq 0$ can occur, however only if the
system is polar.  For a symmetric system with $W_i(x)=W_i(-x)$ and
$\omega_i(x)=\omega_i(-x)$ the steady state distributions are also
symmetric $P_i(x)=P_i(-x)$. Since $\partial_x W_i$ is antisymmetric in
this case, $v=0$ by symmetry according to Eq. (\ref{eq:vss}). On the
other hand, $r$ is in general nonzero in this case (the functions
$\alpha_i$ are symmetric).  For spontaneous motion to occur two
requirements have to be fulfilled: detailed balance of the transition
has to be broken, which corresponds to $\Delta\mu\neq 0$ and the
system must have polar symmetry. In the case of motor proteins the
polar filaments play this role.

In the presence of an external force $f_{\rm ext}$, the system can
perform mechanical work, i.e. it operates as a motor. The work
performed per unit time against the external force is
\begin{equation}
{\cal W} =-f_{\rm ext} v
\end{equation}
while the chemical energy consumed per unit time is given by
\begin{equation}
{\cal Q}=r \Delta\mu \quad .
\end{equation}
We can therefore define the efficiency of energy transduction
\begin{equation}
\eta=-\frac{f_{\rm ext} v}{r \Delta\mu} \quad
\end{equation}
This quantity is useful for forces applied opposite to the direction
of motion where $0\leq \eta \leq 1$. Note, that this definition relies
on the fact that a bulk solution exists which plays the role of a
thermodynamic reservoir and allows to define the chemical potential
difference of fuel and products. In situations where reservoirs are
small the efficiency would be more difficult to define.

For a discussion of physical aspects of motion, it is
useful to write
\begin{equation}
\omega_1(x) = \omega_2(x)(\Omega(x)+e^{-\Delta W/k_B T}) \quad ,
\label{eq:Omega_def}
\end{equation}
where
\begin{equation} 
\Omega(x)=e^{-\Delta W/k_B T}(e^{\Delta\mu/k_BT}-1)\alpha/(\alpha+\beta) \label{eq:Omega}
\end{equation}
measures locally the rate of transitions violating detailed balance.
Using the dependence of the chemical potential on
particle concentration, $\mu_i=\mu^0_i+k_B T \ln C_i$, we observe
that 
\begin{equation}
\Omega \sim \left(\frac{C_{ATP}}{C_{ADP}C_P} - k^0\right) \quad ,\label{eq:simOmega}
\end{equation}
where $k^0=e^{(\mu^0_{ATP}-\mu^0_{ADP}-\mu^0_P)/k_BT}$ is the
equilibrium constant of the hydrolysis reaction. $\Omega$ therefore is
a direct measure of the distance from chemical equilibrium.  From
Eqns. (\ref{eq:Omega}) and (\ref{eq:simOmega}) we find
\begin{equation}
\Omega\sim \left \{
\begin{array}{l l}
\Delta\mu & \hbox{\rm for}\quad \Delta\mu/k_B T \ll 1 \\
C_{ATP}/C_{ADP}C_P & \hbox{\rm for}\quad \Delta\mu/k_B T \gg 1 \\
\end{array}
\right .
\end{equation}
For our discussion of the two-state model it is useful to characterize
the system by the functions $\omega_2$ and $\Omega$ instead of
$\alpha$ and $\beta$ which allows us to discuss motion generation
without the need to introduce chemistry.  This choice is more general
and can be used also for cases where transitions between states are
triggered by other processes than chemistry such as in artificially
constructed systems \cite{rous94,fauc95a,gorr96}.

The properties of this two-state models have been discussed in
Refs. \cite{juli97b,ajda92,chau94,chau95}. Calculating the average
velocity $v$ as a function of the externally applied force $f_{\rm
ext}$ typically leads to a behavior which is well approximated by a
linear dependence 
\begin{equation}
v\simeq v_0(1 -(f_{\rm ext}/f_s)) \label{eq:lin}
\end{equation} 
for many different choices of the potential shapes and the transition
rates.  Here $v_0$ is the spontaneous velocity at zero force $f_{\rm
ext}=0$ and $f_s$ the stalling force, i.e. the force for which the
system stops moving. Deviations from this linear behavior mainly occur
for forces larger than the stalling force $|f_{\rm ext}|>|f_s|$ or for
forces parallel to its natural direction of motion $f_{\rm ext}/f_s<0$.

The observed force-velocity curves for kinesin motors show
an almost linear behavior which can be characterized by $v_0$ and
$f_s$ defined in (\ref{eq:lin}). While $f_s \simeq 5 pN$ does not
depend much on experimental conditions, the no-load velocity $v_0$ depends
on ATP concentration and attached viscous loads and is of the order of
$1 \mu/s$ or smaller \cite{hunt94,svob94}. 

The orders of magnitude observed for kinesin can be reproduced by the
two-state model.  Using e.g. a choice of potentials as shown in
Fig. \ref{f:kinesin} with transitions localized at the potential
minimum, the stall force is approximatively given by the potential
slope. Choosing a potential amplitude of $U\simeq 10 k_B T$ which is
set by the available chemical energy of $\Delta\mu\simeq 10-15 k_BT$
and a period of $l\simeq 8nm$ of microtubules, this force is $f_s
\simeq U/l = 5$pN consistent with the observed value. The spontaneous
velocity of the two-state model can be estimated by $v_0 \simeq
l/(t_c+t_s)$, where $t_c$ is the time of the chemical steps and
$t_s\simeq l^2\eta/U$ is the sliding time in the potential. Therefore,
the observation of $v_0$ does not fix both the chemical rate and the
value of $\eta$ corresponding to protein friction.  One estimate for
the unknown friction $\eta$ is to assume a hydrodynamic friction with
a viscosity $\eta_{\rm vis}$ a factor $10^2-10^3$ larger than the one
of water, suggesting $\eta\sim
\eta_{\rm vis} l \sim (1-10) 10^{-8}$kg/s. 
This value takes into account that the dissipation of proteins should
be better represented by the viscous behavior of macromolecular
solutions. A different approach is to assume that for large ATP
concentration $t_c\simeq
\omega_2^{-1}\Omega^{-1}$ is negligible and the friction $\eta$
determines the sliding velocity $v_0\sim U/l\eta$. Estimating the
maximal velocity to be $v_{\rm max} \sim 10^{-5}$m/s, we find $\eta\sim
10^{-7}$ which can be seen as an upper bound since chemical steps
which in general also contribute to friction are neglected.

A key parameter characterizing the conditions of operation of the
two-state model is the the dimensionless value $U/\xi l^2
\Omega\omega_2$ which compares the typical chemical transition time
with sliding times in the potential slope. With the arguments given
above we estimate $U/\xi l^2 \Omega\omega_2 \simeq 0.1-1$, where we
have used $\Omega \omega_2\simeq 10^{3}$s$^{-1}$ which is a typical
transition rate \cite{gilb95}. However, different values are also
consistent with the observed force-velocity relation as the
spontaneous velocity $v_0$ is determined by the longest of the two
time scales mentioned above.  Additional information such as velocity
fluctuations would be required to determine this value from
experimental observations and to fix the orders of magnitude of all
parameters of the model. 

The two state model is consistent with the observed behaviors for
biological motor molecules and reproduces typical velocities and
forces and the force velocity relation.  Other types of models which
use different representations of states and transitions have also be
used to consistently describe the force-velocity relation of kinesin
\cite{pesk95,duke96,dere96}.
\begin{figure}
\centerline{\psfig{figure=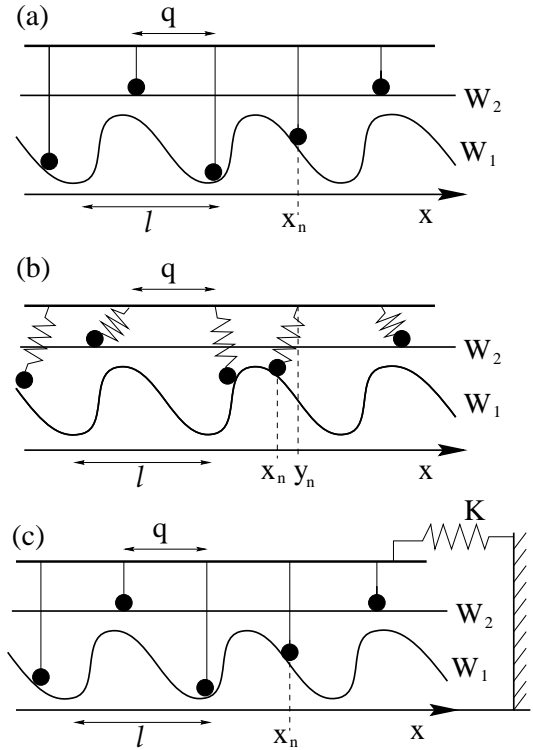,width=7cm}}
\caption{Many motor system as rigidly coupled two-state models. (a)
rigid coupling (b) elastic coupling to rigid backbone (c) elastic
coupling to environment.}
\label{f:coll_mod}
\end{figure}

\section{Collective Effects: Dynamic Instabilities}

In many biological situations, motor molecules and filaments do not
operate as isolated enzymes but many motors are integrated in larger
structures. Typical examples are actin/myosin in muscles and
dyneins/microtubules in flagella and cilia.  Furthermore, the presence
on motor molecules in the cytoskeleton leads to complex physical
properties of these systems on large scales \cite{nede97}.

The most prominent many-motor system is actin/myosin in muscles.  In
this case, myosin molecules are attached together by their tails to
form a linear filament. Myosin filaments and actin filaments are
arranged in parallel in a highly organized fashion.  In the presence
of ATP they slide with respect to each other which macroscopically
leads to muscle contraction.  Experimental in vitro ``motility
assays'' can be used to study myosin function in an artificial
environment. In these systems, myosin molecules are attached to a
solid substrate using specific antibodies. Actin filaments in solution
adsorb to the myosin coated surface and start to move in presence of
ATP as a result of the action of myosin motors \cite{spud90,hara87,wink95,rive98}.

Another example are cilia and flagella which are elastic linear
extensions of many cells which generate a beating motion used to
propel the cell within a solvent or to move the solvent. A flagellum
typically contains 9 pairs of microtubules, each pair coupled by a
large number of dynein motors and other proteins which serve as
structural elements. The motors create forces that lead to the bending
and motion of the flagella. Interestingly, these motors are used to
generate oscillating motion \cite{albe94}.

These systems demonstrate that the behavior of systems containing
motors can on larger scales have new and different types of behaviors
than the one observed for individual motor molecules.  As a first
approach to discuss the behavior of many motor systems, we generalize
the two-state model to describe a large number of coupled motors
\cite{juli97b,juli95,juli97}.
A simple modelization of such a situation is sketched in Fig.
\ref{f:coll_mod} (a). Many motors which all are described by a two-state model are attached along a linear backbone with constant spacing $s$. Assuming
that the spacing between motors is fixed implies that the backbone is
rigid. Within this assumption all motors have the same velocity. In
practical cases, elastic properties of the filaments and of a backbone
coupling the motors can become important. For example in the case of
muscles, the passive elastic behavior of proteins such as titin can
play the role of elastically coupling motor-filament systems to their
environment \cite{albe94}.  Similarly, in flagella, bending elasticity of the
filaments is essential to allow for the generation of beating motion.
Fig. \ref{f:coll_mod} (b) and (c) sketches two simple ways to
incorporate the effects of material elasticity in the modelization.

\subsection{Mean field limit}

Motors coupled via a rigid backbone allow to illustrate the appearance
of collective effects. For a system of $N$ elements with two states,
moving along a periodic structure, we can introduce the distribution
function $p(x,\sigma_1,...,\sigma_N)$ for finding the particles
$i=1..N$ in states $\sigma_i=1,2$ with particle $i$ at position $x+si$
along a linear coordinate. This system thus becomes an effective
$2^N$-state system described by $2^N$ equations
\begin{eqnarray}
\partial_t p(x,\sigma_1,..,\sigma_N)&+&
\partial_x j(x,\sigma_1,..,\sigma_N)= -\sum_{i=1}^N \omega_{\sigma_i}(x+is)
p(x,\sigma_1,..,\sigma_N) \nonumber \\
&+& \sum_{i=1}^N \omega_{\bar\sigma_i}(x+is) p(x,\sigma_1,..,\bar\sigma_i,..,\sigma_N)
\end{eqnarray}
Here, $\omega_\sigma(x)=\omega_\sigma(x+l)$ are the individual
transition rates defined in the previous sections and the bar denotes
the opposite state, i.e, $\bar1=2, \bar2=1$. The currents are given by
\begin{equation}
j(x,\sigma_1,..,\sigma_N)=\frac{1}{\eta N}\left [
-k_BT \partial_x p -N p\partial_x w(x,\sigma_1,..,\sigma_N)-N pf_{\rm ext}\right ]
\quad ,
\end{equation}
with the potential
\begin{equation}
w(x,\sigma_1,..,\sigma_N)=\frac{1}{N}\sum_{i=1}^N W_{\sigma_i}(x+is) \quad ,
\end{equation}
defined as a sum of individual particle potentials.  Here, $\eta N$
denotes the total friction which is assumed to scale linearly with $N$
and $f_{\rm ext}$ is the externally applied force per motor.

In order to reduce the number of equations and to obtain a tractable
description, we introduce the average density of particles found at
position $\xi=x\;\hbox{\rm mod}\; l$, $0\leq\xi\leq l$ relative to the
potential period:
\begin{equation}
P_k(\xi)=<\rho_k(\xi)> \quad ,
\end{equation}
where 
\begin{equation}
\rho_k(\xi)=\frac{1}{N}\sum_{i=1}^N \delta_{k,\sigma_i}\delta(x+is-\xi) \quad .
\end{equation}
Here, we have introduced the notation
\begin{equation}
<a>=\lim_{m\rightarrow\infty}\frac{1}{2m}\int_{-ml}^{ml} dx \sum_{\sigma_1..\sigma_N} a(x,\sigma_1,..,\sigma_N)p(x,\sigma_1,..,\sigma_N) \quad ,
\end{equation}
for averages over the distribution $p$ which we assume to be periodic,
$p(x,\sigma_1,..,\sigma_N)=p(x+l,\sigma_1,..,\sigma_N)$ normalized
over one period $l$, $<1>=1$. The densities $P_k(\xi)$ satisfy the
normalization condition Eq. (\ref{eq:norm2s}) and behave like a single
particle two-state model
\begin{eqnarray}
\partial_t P_1 + \partial_\xi J_1 &=& -\omega_1 P_1 + \omega_2 P_2 \nonumber \\
\partial_t P_2 + \partial_\xi J_2 &=&\omega_1 P_1 - \omega_2 P_2 \quad ,
\end{eqnarray}
however with the currents
\begin{equation}
J_k(\xi)=-\frac{k_BT}{N\eta}\partial_\xi P_k-
<\rho_k(\xi)v>
\quad ,\label{eq:Jk}
\end{equation}
where $v(x,\sigma_1,..,\sigma_N)=
-[\partial_x w(x,\sigma_1,..,\sigma_N)+f_{\rm ext}]/\eta$.
From now on, we consider a large number $N$ of motors
and we assume that the period of motors is incommensurate with the
potential period, $s/l$ irrational. In this case,
the total particle distribution function $P_1+P_2=<\sum_{i=1}^N
\delta(\xi-x-is)/N>$ becomes homogeneous:
\begin{equation}
P_1(\xi)+P_2(\xi)=\frac{1}{l} +O(1/N) \quad ,
\end{equation}
and we can approximate $<\rho_k(\xi)v>=<\rho_k(\xi)><v> +O(1/N)$.
Ignoring terms of order $1/N$ including the diffusive term in
Eq. (\ref{eq:Jk}) the currents simplify to
\begin{equation}
J_k(\xi)= v P_k(\xi) \quad ,
\end{equation}
where
\begin{equation}
v=<v>= \frac{1}{\eta}\left[-\int_0^l d\xi(P_1\partial_\xi W_1+P_2\partial_\xi W_2)
+ f_{\rm ext}\right]\quad .
\end{equation}
We have found a simple mean-field theory which is very useful to
explore the properties of a many-motor system.  Ignoring all
corrections in $1/N$, we finally obtain
\begin{eqnarray}
\partial_t P_1+v\partial_\xi P_1=-(\omega_1+\omega_2)P_1 +\frac{\omega_2}{l} 
\label{eq:coll_mod_P}\\
v=\frac{1}{\eta}\left[\int_0^l d\xi P_1\partial_\xi \Delta W +f_{\rm ext}
\right] \label{eq:coll_mod_v}
\end{eqnarray}
which describes the time-evolution of $P_1(\xi)=1/l-P_2(\xi)$.

\subsection{Steady states}

First, we look at the properties of steady state solutions with
$\partial_t P_1=0$ and constant velocity which obey \cite{juli95}
\begin{equation}
v\partial_\xi P_1=-(\omega_1+\omega_2)P_1 +\omega_2/l \quad ,\label{eq:collss}
\end{equation}
where $v$ is a parameter.  Eq. (\ref{eq:collss}) can be solved
analytically for simple choices of the transition rates. A more
general approach is a power expansion of the steady state in the
velocity
\begin{equation}
P_1(\xi)=\sum_{n=0}^\infty P_1^{(n)}(\xi) v^n \quad,
\end{equation}
where the $P_1^{(n)}$ satisfy the recursion relation
$P_1^{(n)}=-(\partial_\xi P_1^{(n-1)})/(\omega_1+\omega_2)$ with
$P_1^{(0)}=\omega_2/(\omega_1+\omega_2)l$. Inserting this solution
in Eq. (\ref{eq:coll_mod_v}), we obtain a force-velocity relation
$f_{\rm ext}=f(v)$, with
\begin{equation}
f(v)=F^{(0)}_\Omega+(\eta+F^{(1)}_\Omega)v + 
\sum_{n=2}^\infty F^{(n)}_\Omega v^n \quad , \label{eq:fv}
\end{equation}
where
\begin{equation}
F^{(n)}_\Omega=-\int_0^l d\xi P_1^{(n)} \Delta W' \quad ,
\end{equation}
and the prime denotes a derivative with respect to $\xi$.  Using the
definition Eq. (\ref{eq:Omega_def}) for the transition rates, 
we find the spontaneous force
\begin{equation}
F^{(0)}_\Omega=-\int_0^l d\xi \frac{\Delta W'}{\Omega+e^{-\Delta W/k_BT}} 
\quad .
\end{equation}
This force is zero for a system which is not chemically driven, i.e.
$\Omega=0$ or if the system is symmetric.  Similarly,
\begin{equation}
F^{(1)}_{\Omega}=\int_0^l d\xi\frac{(\Delta W')^2 e^{-\Delta W/k_BT}
-\Omega'\Delta W'}{l\omega_2(\Omega+1+e^{-\Delta W/k_BT})^3}
\end{equation}
is an effective friction coefficient which results from chemical
transitions. For a passive system, this friction must be positive,
$F^{(1)}_{\Omega=0}>0$, which is indeed the case. For $\Omega\neq 0$,
however, $f^{(1)}$ can become negative if $\Omega'\Delta W'$ is
positive. This criterion implies that the function $\Omega(\xi)$ has
maxima and minima at the same positions as $\Delta W$. For $W_1$
periodic and $W_2$ constant this suggests e.g. placing the maxima of
$\Omega$ at the positions of minimal $W_1$ which is the idea of
localized excitations.

\begin{figure}
\centerline{\psfig{figure=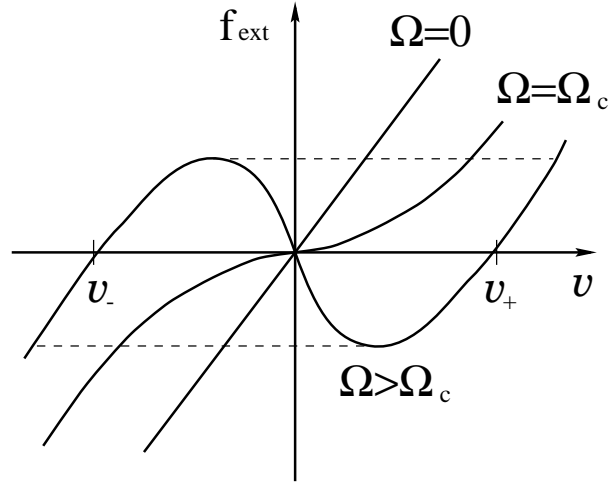,width=8cm}}
\caption{Relation between velocity $v$ and externally applied force
$f_{\rm ext}$ for a symmetric system. For $\Omega<\Omega_c$ the system
is passive. A dynamic transition with spontaneous symmetry-breaking
occurs for $\Omega=\Omega_c$. For $\Omega>\Omega_c$, the system is
actively moving with velocity $v_\pm$. The hysteresis is indicated
by a broken line.}
\label{f:sym_sys}
\end{figure}

The fact that $F_\Omega^{(1)}$ can become negative suggests the
possibility of an instability in the system.  This can be
discussed most easily by first considering a symmetric system with
symmetric potentials and transition rates. In this case, the system is
invariant under $(v,f_{\rm ext})\rightarrow (-v,-f_{\rm ext})$,
indicating that all even coefficients $f_\Omega^{(2n)}=0$ vanish by
symmetry.  In this case,
\begin{equation}
f(v)=(\eta+F_\Omega^{(1)})v + F_\Omega^{(3)} v^3 + O(v^5) \quad .
\end{equation}
If $F^{(1)}_\Omega$ is negative and decreases with increasing
amplitude of $\Omega$, an instability occurs as soon as $\Omega=\Omega_c$
for which
\begin{equation}
F^{(1)}_{\Omega_c}=-\eta \quad .
\end{equation}
For $\Omega>\Omega_c$, the function $f(v)$ has a maximum, a
minimum and a velocity interval where the effective friction 
\begin{equation}
\eta_{\rm eff}=\frac{\partial f(v)}{\partial v}
\end{equation}
becomes negative, see Fig. \ref{f:sym_sys}. If no
external force is applied, the system does not move for
$\Omega<\Omega_c$ but two moving solutions bifurcate for
$\Omega>\Omega_c$ with opposite velocity while the non-moving solution
becomes unstable as revealed by a linear stability analysis. The
system therefore at $\Omega=\Omega_c$ undergoes a bifurcation with
spontaneous symmetry breaking which allows for the existence of
spontaneous motion in the symmetric case.

If an external force is applied, the system can work against this
force until the instability is reached at the minimum of the force
velocity relation. Further increase of the force leads to a
discontinuous change of the velocity. The system therefore shows the
signature of a first-order transition with hysteretic behavior and for
$f_{\rm ext}=0$ bistability where the selection of one of the stable
states depends on the history of the system.

These arguments can be extended to the general case of a system with
polar symmetry. In this case, all coefficients and in particular the
spontaneous force $F^{(0)}_\Omega$ are nonzero. Dynamical transitions
and instabilities do still exist but now occur typically for nonzero
velocity and in an asymmetric way. A transition occurs if for
increasing $\Omega$ a point with $\eta_{\rm eff}=0$ exists for a
critical velocity $v_c$ and for a critical force $f_c$.  For
$\Omega>\Omega_c$, a finite region with $\eta_{\rm eff}=0$ emerges and
the function $f(v)$ has a maximum and a minimum. In this case,
discontinuous changes of the velocity occur as a function of applied
force as soon as the force reaches the extremal values where the
system becomes unstable and the transition shows a hysteresis. A
behavior with the characteristic as predicted by this theoretical
analysis has been observed in motility assay experiments where
electric fields are used to exert an external force \cite{rive98}.

\subsection{Spontaneous oscillations}

\begin{figure}
\centerline{\psfig{figure=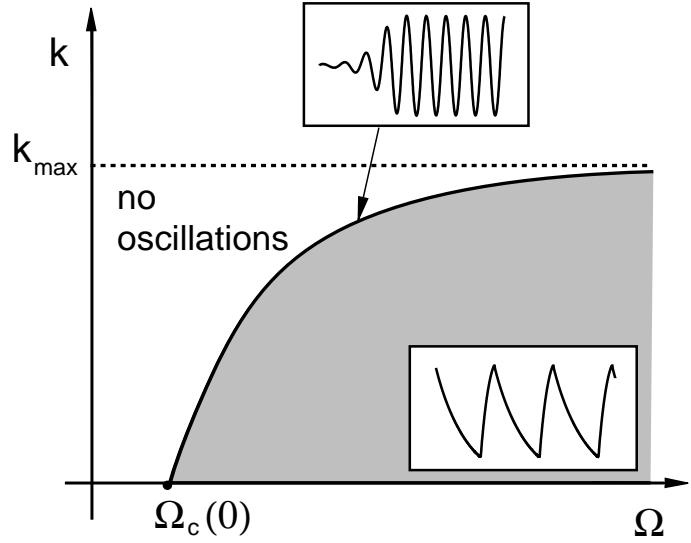,width=9cm}}
\caption{Phase diagram for spontaneous oscillations as a function
of the excitation activity $\Omega$ and the modulus per motor $k$ of
an external elastic element. The resting state becomes unstable with
respect to oscillations along a line with $\Omega=\Omega_c(k)$. The
insets show typical examples for oscillatory motion (displacement
versus time): onset of sinusoidal motion at the instability and
nonlinear oscillations with cusp-like shape for small $k$ and large
$\Omega$.}
\label{f:so}
\end{figure}

Another important consequence of the collectivity is the possibility
to generate oscillatory motion if the system is acting together with
elastic elements, see Fig. \ref{f:coll_mod} (c). This corresponds to
adding an elastic element to the force balance
Eq. (\ref{eq:coll_mod_v}), which leads to
\begin{equation}
v=\frac{1}{\eta}\left[\int_0^l d\xi P_1 \Delta W' +f_{\rm ext} -k x 
\right]\quad , \label{eq:vosc}
\end{equation}
where $K=k N$ is the modulus of an external elastic element, $k$,
the modulus per motor and
\begin{equation}
x(t)=\int_0^t dt' v(t') \quad,
\end{equation}
is the total displacement of the system. First, we assume that the
modulus $k$ is small. In this case, the force varies slowly and we can
use an adiabatic approximation assuming that at any given time the
system is in one of the steady states with
\begin{equation}
f_{\rm ext}-k x = f(v) \quad ,
\end{equation}
where $f(v)$ is the force-velocity relationship introduced in Eq.
(\ref{eq:fv}). The change in velocity can be expressed as
\begin{equation}
\dot v = -\frac{K v}{\eta_{\rm eff}(v)} \quad . \label{eq:osc_adia}
\end{equation}
This equation demonstrates that we expect an oscillatory instability
to occur in this adiabatic limit, exactly if
\begin{equation}
\eta+F^{(1)}_\Omega < 0 \quad. \label{eq:instosc}
\end{equation}
In this case, the non-moving state with $v=0$ is unstable. As soon as
the system starts moving an elastic force is created which opposes
motion and leads to a decrease in velocity according to
Eq. (\ref{eq:osc_adia}). As soon as an instability is reached with
$\eta_{\rm eff}=0$, the velocity changes discontinuously in the
opposite direction and the process is repeated. This scenario leads to
characteristic relaxation oscillations of the position with a sawtooth
like shape. For a symmetric system, the forward and backward parts of
this process are identical and oscillations are symmetric with respect
to $t\rightarrow -t$. In the asymmetric case, however, the asymmetry
of the system is reflected in the shape of the oscillations, see
inset of Fig. \ref{f:so}.

This adiabatic approximation is valid as long as the characteristic
time $\eta_{\rm eff}/k$ is much longer than the transition times, i.e.
$\eta_{\rm eff}/k\gg \omega_1^{-1}+\omega_2^{-1}$. For large $k$ and
in particular close to the discontinuous transition with $\eta_{\rm
eff}=0$, the full dynamic equations have to be taken into account.
The onset of oscillatory motion can be determined by a linear
stability analysis of the non-moving state with $v=0$, and
$x=x_0=-f^{(0)}_\Omega/k$.  For a small deviation from this initial
state we write the ansatz
\begin{eqnarray}
P_1(\xi,t) & = & R(\xi) + p(\xi) e^{s t} \\
v(t) & = & u e^{s t} \quad ,
\end{eqnarray}
where $R\equiv P_1^{(0)}$ and $s=i\omega+\tau$ is complex. Here,
$\omega$ is an oscillation frequency and $\tau$ a relaxation time.
Inserting these expressions in Eq. (\ref{eq:coll_mod_P}), we find to
linear order
\begin{equation}
p(\xi) = -u \frac{R'}{\alpha+s} \quad ,
\end{equation}
where we have defined $\alpha(x)=\omega_1(x)+\omega_2(x)$.
Possible values of $s$ are solutions to
\begin{equation}
\eta + \frac{k}{s} =- \int_0^l d\xi \frac{R' \Delta W'}{\alpha
+s}
\quad , \label{eq:ses}
\end{equation}
As long as $\tau <0$, the initial state is locally stable, it becomes
unstable for $\tau=0$ towards oscillatory motion with frequency
$\omega$. This instability is determined by setting $s=i\omega_c$. The
frequency $\Omega_c$ at the instability obeys
\begin{eqnarray}
\eta & = & -\int_0^l d\xi \frac{\alpha}{\alpha^2 +\omega_c^2} R'
\Delta W' \\
k & = & \int_0^l d\xi \frac{\omega_c^2}{\alpha^2 +\omega_c^2}R'
\Delta W' \label{eq:keta} \quad .
\end{eqnarray}
For $k=0$, we recover the criterion (\ref{eq:instosc}) with
$\omega_c=0$.  The frequency at the instability increases as
$\omega_c\sim k^{1/2}$ for increasing values of $k$. However, for
$k>k_{\rm max}$ with
\begin{equation}
k_{max}\leq -\int_0^l R \Delta W'' \quad ,
\end{equation}
no instability occurs for arbitrarily large $\Omega$ see Fig.
\ref{f:so}.  Note, that
$k_{\rm max}$ is of the order of magnitude of the average negative
curvature of the potential $\Delta W$ and can be roughly estimated by
$k_{\rm max}\simeq U/l^2$, where $U$ is the potential period. Note,
that since $k$ a modulus per motor, a motor collection of $N$ motors
can therefore undergo oscillatory motion if working against an elastic
element with rigidity smaller than $K<k_{\rm max} N$.

The nonlinear relation between force and velocity of steady states can
be generalized to time-dependent external forces.  Assuming that a
limit cycle exists which is characterized by the Fourier
representations
\begin{eqnarray}
f_{\rm ext}(t)& =& \sum_{n=-\infty}^\infty f_n e^{i \omega t} \nonumber\\
v(t)& =& \sum_{n=-\infty}^\infty v_n e^{i \omega t} \nonumber\\
\end{eqnarray}
a generalization of the expansion (\ref{eq:fv}) is given by
\begin{equation}
f_n = F^{(0)}_n+ F^{(1)}_{n,l} v_k + F^{(2)}_{n,lm}v_l v_m + F^{(3)}_{n,lmo}
v_l v_m v_o + O(v^4) \quad . \label{eq:exp_osc}
\end{equation}
The spontaneous force $F^{(0)}_n=F^{(0)} \delta_{n0}$ is nonvanishing only
for the time independent mode $n=0$. The linear response coefficient
\begin{equation}
F^{(1)}_{n,l}=\delta_{nl}\left[\eta + \frac{k}{i\omega l} +\int_0^l d\xi
\frac{R'\Delta W'}{\alpha + i\omega l} \right ]
\end{equation}
vanishes at the instability. The real and the imaginary part of
$F^{(1)}_{n,m}$ corresponds to an effective friction $\eta_{\rm eff}$
and to an effective elastic modulus $k_{\rm eff}/i\omega$,
respectively. Because of the non equilibrium nature of the systems
both of these effective linear response coefficients can become
negative which is impossible in an equilibrium situation.  

Higher order coefficients can be obtained in a systematic way
\cite{juli97}. The expansion (\ref{eq:exp_osc}) characterizes the
history-dependent response of the system to an external force which is
periodic in time for small velocity amplitudes.  As a result of time
translation symmetry, the coefficients obey $F^{(n)}_{m,l_1..l_n}\sim
\delta_{m,l_1+..+l_n}$. Spontaneous oscillations are solutions
to Eq. (\ref{eq:exp_osc}) with $f_n=0$ for all $n$ which are well
described near the instability. Far away from the instability
nonlinearities become dominant. In the regime of small $k$, they
are well captured by the adiabatic approximation given by
Eq. (\ref{eq:osc_adia}).

\section{Discussion and outlook}

In the previous sections, we have described a generic framework to
model the force and motion generation of molecular motors. The
coupling of a chemical reaction to internal conformations and spatial
degrees of freedom of a motor enzyme is described by an overdamped
stochastic dynamics. Reducing the description to a simple two state
model has the advantage to be sufficiently simple for a theoretical
analysis but still flexible to allow for a description of conditions
of operation similar to those of motor molecules such as myosin or
kinesin.

A quantitative modelization of motor enzymes on the molecular level is
difficult. One the one hand experimental data is limited, on the other
hand simplified models are not designed to describe all details of the
functioning of complex enzymes.  The properties of the two-state
model discussed here depend of the choice of the potential shapes and
on the conformation-dependence of the transition rates described by the
functions $\alpha(x)$ and $\beta(x)$.  These functions
cannot be deduced from the molecular structure and are difficult to
obtain experimentally. This problem is often
referred to as the unknown strain-dependence of transitions
\cite{leib93,duke96}.  The known force-velocity relation
for single kinesin molecules is close to linear \cite{svob94,hunt94}.
Different models which have been used to describe kinesin motion, all
produce almost linear force-velocity relations and describe well the
observed orders of magnitude \cite{pros94,pesk95,duke96,dere96} which
demonstrates that the basic properties of force-generation do not
depend much on the details of a  chosen model.  For these reasons, we
refrain from a full quantitative comparison and focus our discussion
on physical principles.

In biological situations, molecular motors are involved in processes
such as muscle contraction, cell division and flagellar beating which
are complex and involve large numbers of motors together with other
enzymes which control and regulate the function of these systems. A
physical approach to more complex situations is to avoid the
complexity of control systems and to investigate the types of behavior
which can result from the activity of many motors alone.  The
properties of larger-scale active systems can be studied using
concepts of statistical physics.  An important advantage of two-state
models in this context is the possibility to start from a simple model
for individual motors and to address the properties of many-motor
systems. From such an approach the possibility of instabilities,
dynamic transitions and spontaneous oscillations follows quite
naturally. We therefore predict such behaviors to occur in biological
situations. Related collective phenomena have recently been shown to
exist also in other models \cite{vand98,reim99} which shows that the
existence of dynamic instabilities in this class of systems is quite
general.  The detailed conditions for which dynamic transitions will
occur in biological systems such as coupled myosins or kinesins should
depend on various parameters such as salt concentrations, temperature,
motor density etc. Our approach does not predict
for which situations these new behaviors can be found but
provides a classification of the possible self-organized behaviors of
such systems.

In vitro motility assays which use purified motors and filaments in an
artificial environment are ideal systems to test these ideas and to
study motor action in the absence of further biochemical regulatory
systems.  Motors grafted to a substrate set fluorescently marked
filaments in motion which can easily be observed by standard light
microscopy.  The force-velocity relationship of the actin-myosin
system has recently been determined by constraining filament motion
along one dimension by using linear grooves of $1\mu$m in diameter as
a substrate \cite{rive98,ottln}, see Fig. \ref{f:motexp} (d).
Electric fields $E$ applied parallel to the aligned and negatively
charged actin filaments induce external forces which are homogeneously
distributed along the filaments. The electrophoretic mobility
$\mu=v/E$, where $v$ is the velocity induced by an electric field $E$,
measured for free filaments can be used to obtain an estimate for the
external force per unit length $f\simeq 2\pi
\mu\eta_w/\ln (r/d) E$ acting effectively on the filament. Here, $eta_w$ 
is the viscosity of the solution and $r$ and $d$ are the radius of the
filament and the distance between filament axis and the substrate.
With this argument, we find that a field of $E\simeq 10^2$V/m induces a
force per length of $f\simeq 1 pN/\mu$m which is the order of
magnitude of stall forces of an actin-myosin system \cite{rive98}. 

In addition to providing the first force-velocity relations for a
purified actin-myosin system, these experiments have revealed
unexpected behavior for high myosin densities close to stall
conditions. With increasing force applied against the direction of
motion, filaments slow down, however before stalling completely an
abrupt change in velocity to reverse motion is observed.  Since a
histogram of the velocity distribution reveals two distinct maxima for
different coexisting velocities and discontinuous changes of the
velocity occur with a hysteretic behavior, a natural interpretation of
this observation is a dynamic first order transition of the type
discussed in the previous sections. However, experiments of this type
are difficult and measured values have significant uncertainties; more
data is necessary to clearly prove the existence of the transition.
Interestingly, macroscopic measurements of the behavior of muscles
show near stalling conditions an unexpected rapid drop in velocity and
a contracting state which is difficult to stabilize experimentally by
feedback systems \cite{edma88}. This behavior could result from the averaged
mechanical properties of many different filaments which individually
show a transition of the type observed in motility assays. 

In recent years, an increasing number of situation have been observed
for which the concepts of dynamic instabilities in many-motor systems
could be relevant.  It is well known that the flight muscles of many
insects generate oscillations themselves. These so-called asynchronous
muscles show oscillatory behavior which is not triggered by a periodic
nerve-signal \cite{prin77}.  These systems are complete
muscles for which the role of biochemical control systems could be
important. Recently, it has become possible to study the active
mechanical properties of single myofibrils, i.e. contractile fibers
within muscle cells, by using microneedles. Ordinary muscle myofibrils
which under normal conditions simply contract, show spontaneous
oscillations if put in specific conditions such as increased ADP
concentration. Both tension- and length- oscillations were observed
\cite{yasu96}. It was demonstrated that such oscillations continue to
exist if regulatory enzymes bound to the actin-myosin motors were
removed \cite{fuji98}. These observations demonstrate the general 
possibility of oscillations in many muscles. The facts that in
muscular structures, which couple contractile elements elastically,
oscillations are typically observed while the motility assay which
lacks elastic elements reveals a first-order transition is fully
consistent with the collective effects discussed here.

Another important type of oscillating many-motor systems are cilia
and flagella. These are hair-like appendages of many cells which are
used for self-propulsion and to stir the surrounding fluid. They share
the characteristic architecture of their core structure, the axoneme,
a common structural motive that was developed early in evolution. It
is characterized by nine parallel pairs of microtubules which are
arranged in a circular fashion together with a large number of dynein
molecular motors. In the presence of ATP, the dyneins attached to the
microtubules generate relative forces while acting on neighboring
microtubules. The resulting internal stresses induce the bending and
wave-like motion of the axoneme \cite{albe94}.
A simple two-dimensional model of filaments driven internally by molecular
motors can be used to
demonstrate that a dynamic instability of the many-motor system most
naturally generates oscillating motion and wave-like propagating
shapes which can be used for self-propulsion \cite{cama99}.  This idea
that simple patterns of motion are induced via a dynamic instability
is supported by the observations that flagellar dyneins are able to
generate oscillatory motion on microtubules \cite{shin98}, and that
isolated and de-membranated flagella in solution containing ATP above
a threshold concentration swim with a simple wave-like motion
\cite{gibb75}.  This suggests, that basic
wave-like patterns of flagellar motion are generated by a
self-organized motor-induced mechanism and that sophisticated
regulatory systems could have evolved at a later time to fine tune the
system and to generate new patterns of motion.

The author thanks A. Ajdari, M. Badoual, M. Bornens, S. Camalet,
P. Chaikin, R. Everaers, A. Maggs, A. Ott, A. Parmeggiani, J. Prost,
D. Riveline and C. Wiggins for stimulating interactions.


\begin{thebibliography}{99}
\bibitem{albe94} Alberts B.,
Bray D., Lewis J., Raff M., Roberts K. and Watson J.D. (1994): The molecular
biology of the cell, (Garland, New York).
\bibitem{huxl57} Huxley A.F. (1957): Prog.
Biophys. {\bf 7}, 255.
\bibitem{huxl69} Huxley H.E. (1969): Science {\bf 164}, 1365.
\bibitem{krei93} Kreis T., and Vale R. (1993): {\em Cytoskeletal and Motor Proteins}, (Oxford University Press, New York).
\bibitem{huxl71} Huxley A.F., and
Simmons R.M. (1971): Nature {\bf 233}, 533.
\bibitem{hill74} Hill T.L. (1974): Prog. Biophys. Mol. Biol.
{\bf 28}, 267.
\bibitem{spud90} Spudich J.A. (1990):, Nature {\bf 348}, 284.
\bibitem{juli97b} J\"ulicher F., Ajdari, A. and Prost J. (1997):
Rev. Mod. Phys. {\bf 69} 1269.
\bibitem{kron86} Kron S.J. and Spudich J.A. (1986): 
Proc. Natl. Acad. Sci. USA {\bf 83}, 6272.
\bibitem{hara87} Harada Y., Noguchi A., Kishino A.,
and Yanagida T. (1987): Nature {\bf 326}, 805. 
\bibitem{ishi91} Ishijima A., Doi T., Sakurada K., and Yanagida T. (1991):
Nature {\bf 352} 301.
\bibitem{wink95} Winkelmann D.A., Bourdieu L., Ott A.,
Kinose F., and Libchaber A. (1995): Biophys. J. {\bf 68} 2444.
\bibitem{fine94} Finer J.T., Simmons R.M., and Spudich J.A. (1994):
Nature {\bf 368}, 113.
\bibitem{bloc98} Block S.M. (1998): J. Cell. Biol. {\bf 140}, 1281.
\bibitem{svob93} Svoboda K., Schmidt C.F., Schnapp B.J., and Block S.M. 
(1993): Nature {\bf 365}, 721.
\bibitem{schn97} Schnitzer M. and Block S.M. (1997): Nature {\bf 388}
386.
\bibitem{hunt94} Hunt A.J., Gittes F., and Howard J. (1994): 
Biophys. J. {\bf 67}, 766.
\bibitem{rive98} Riveline D., Ott A., J\"ulicher F., Winkelmann D.A., 
Cardoso O., Lacap\`ere J.-J., Magnusdottir S., Viovy J.-L.,
Gorre-Talini L. and Prost J. (1998): Eur. Biophys. J. {\bf 27} 403.
\bibitem{mand99} Mandelkow E. and Hoenger A (1999): Current Opinion in
Cell Biology {\bf 11}, 34.
\bibitem{okad99} Okada Y. and Hirokawa N. (1999): Science {\bf 283}, 1152.
\bibitem{feyn66} Feynman R.P., Leighton R.B., and Sands M.
(1966): {\em The Feynman Lectures on Physics} (Addison-Wesley, Reading MA),
Vol. I, Chap. 46.
\bibitem{buet87} B\"uttiker, M. (1987): Z. Phys. B {\bf 68}, 161.
\bibitem{land88} Landauer  R. (1988): J. Stat. Phys. {\bf 53}, 233.
\bibitem{ajda92} Ajdari A., and Prost J. (1992): C.R. Acad. Sci. Paris II,
{\bf 315}, 1635.
\bibitem{magn93} Magnasco M.O. (1993): Phys. Rev. Lett. {\bf 71} 1477 (1993).
\bibitem{pros94} Prost J., Chauwin J.F., Peliti L., and Ajdari A. (1994): Phys. Rev. Lett. {\bf 72}, 2652.
\bibitem{pesk94} Peskin C.S., Ermentrout G.B., and Oster G.F. (1994):
{\em Cell Mechanics and Cellular Engineering}, V.Mow et al eds.
(Springer, New-York).
\bibitem{astu94} Astumian R.D., and Bier M. (1994):              
Phys. Rev. Lett. {\bf 72}, 1766.
\bibitem{doer95} Doering C.R. (1995): Nuovo Cimento {\bf 17}, 685.
\bibitem{astu97} Astumian R.D. (1997): Science {\bf 276},
917.
\bibitem{rous94} Rousselet J., Salome L., Ajdari A., and Prost J., 1994, Nature {\bf 370}, 446.
\bibitem{fauc95a} Faucheux L.P., and Libchaber A. (1995):
J. Chem. Soc. Farad. Trans. {\bf 91}, 3163.
\bibitem{gorr96} Gorre L., Ioannidis E., and Silberzan P. (1996):
Europhys. Lett. {\bf 33}, 267.
\bibitem{gorr98} Gorre-Talini L., Spatz J.P., and Silberzan P. (1998):
Chaos {\bf 8} 650.
\bibitem{lymn71} Lymn R.W., and Taylor E.W. (1971):
Biochem. {\bf 10}, 4617.
\bibitem{risk84} Risken H. (1984): {\em The Fokker-Planck Equation}
(Springer, Berlin).
\bibitem{leib93} Leibler, S., and Huse D. (1993): 
J. Cell Biol. {\bf 121}, 1357.
\bibitem{pesk95} Peskin C.S., and Oster G.F. (1995): Biophys. J. {\bf 68},
202.
\bibitem{chau94} Chauwin J.F., Ajdari A., and Prost J. (1994): 
Europhys. Lett. {\bf 27}, 421.
\bibitem{chau95} Chauwin J.F., Ajdari, and Prost J. (1995): 
Europhys. Lett. {\bf 32}, 373.
\bibitem{svob94} Svoboda K, and Block S. (1994): Cell {\bf 77}, 773 (1998).
\bibitem{gilb95} Gilbert S.P., Webb M.R., Brune M., and
Johnson K.A. (1995): Nature {\bf 373} 671.
\bibitem{duke96} Duke T., and Leibler S. (1996): Biophys. J. {\bf 71}, 1235.
\bibitem{dere96} Der\'enyi I., and Vicsek T. (1996): Proc. Nat. Acad. Sci.
{\bf 93}, 6775.
\bibitem{nede97} N\'ed\'elec F.J, Surrey T., Maggs A.C. and
Leibler S. (1997): Nature {\bf 389}, 305.
\bibitem{juli95} J\"ulicher F., and  Prost J. (1995): 
Phys. Rev. Lett. {\bf 75}, 2618.
\bibitem{juli97} J\"ulicher F., and  Prost J. (1997): 
Phys. Rev. Lett. {\bf 78}, 4510.
\bibitem{vand98} Van den Broeck C., Reimann P., Kawai R., and H\"anggi P. (1998): preprint.
\bibitem{reim99} Reimann P., Kawai R., Van den Broek C., 
and H\"anggi P. (1999): Europhys. Lett. {\bf 45}, 545.
\bibitem{ottln} Ott A, contribution in this volume.
\bibitem{edma88} Edman, K.A.P (1988): J. Physiol. {\bf 404}, 301.
\bibitem{prin77} Pringle J.W.S. (1977): in {\em Insect Flight Muscle}, R.T.
Tregear, Ed. , North-Holland, Amsterdam, 177.
\bibitem{yasu96} Yasuda K., Shindo Y. and Ishiwata S. (1996): Biophys. J. {\bf 70}, 1823.
\bibitem{fuji98} Fujita H. and Ishiwata S. (1998):
Biophys. J. {\bf 75}, 1439.
\bibitem{cama99} Camalet S., J\"ulicher F., and Prost J. (1999):
Phys. Rev. Lett. {\bf 82}, 1590.
\bibitem{shin98} Shingyoji C., Higuchi H., Yoshimura M., Katayama E., and 
Yanagida T. (1998): Nature (London) {\bf 393}, 711.
\bibitem{gibb75} Gibbons I.R. (1975): in 
{\em Molecules and Cell Movement}, S. Inou\'e and
R.E. Stephens (Eds.), Raven Press, New York. 
\end{thebibliography}
\end{document}